\documentclass[a4paper,9pt,twocolumn]{article}
\usepackage{graphicx}

\usepackage{subcaption}

\begin{document}

\title{Efficient Calling Conventions for Irregular Architectures}
\author{Philipp K. Krause}





\maketitle

\section*{Abstract}

We empirically evaluated thousands of different C calling conventions for irregular microcontroller architectures, and found potential for improvement over the calling conventions previously used in the Small Device C Compiler (SDCC). The improvements in code size and speed are substantial enough that SDCC made changes to its default calling convention, breaking ABI compatibility.

\section{Introduction}

Calling convention overhead can contribute significantly to code size, power consumption and execution time. Often, calling conventions are chosen per architecture or per operating system early on, and then kept to not break binary compatibility. For many systems, attempts to change the calling convention, and thus the application binary interface (ABI) would meet with substantial opposition from existing users. Thus research on reducing calling convention overhead has focused on choosing a calling convention for some individual functions as a link-time optimization~\cite{Bus2004,Caldwell2017}. That research looked into calling conventions for ARM, a reduced instruction set computer (RISC) architecture. RISC architectures typically have a relatively large number of general-purpose registers that can be used interchangeably. In microcontrollers, sometimes the opposite is found: Accumulator architectures, which have one accumulator, and use memory operands otherwise. For accumulator architectures, compilers often use a number of memory locations as pseudoregisters, so from the compiler perspective the use of pseudoregisters can be handled similarly to registers in RISC architectures. Such architectures are considered compiler-friendly, as traditional graph-coloring register allocators can deal well with them. Calling conventions typically pass return values and many function parameters in registers or pseudo-registers; since the registers or pseudo-registers can be used interchangeably, it doesn't really matter which register is used for which purpose, it only matters if there are enough of them.

However, there are also irregular architectures with a rather small number of registers, with instructions often being available for some subset of registers only, or instruction size and execution time depending on which registers the operands reside in. Recently, register allocators that can deal well with such architectures have been developed~\cite{KrauseRalloc}, which might help make irregular architectures popular again. For such architectures, the choice of an ABI can be a quite interesting problem.

The free~\cite{Stallman2002} Small Device C Compiler~\cite{SDCC} targets various 8-bit architectures common in microcontrollers (µC), as well as some historic 8-bit architectures relevant to the retrocomputing and retrogaming communities. This includes many highly irregular architectures. Typical SDCC users are highly sensitive to code size. Use of precompiled libraries written in C is uncommon (apart from the standard library that ships with SDCC), but some users, as well as downstream projects that bundle SDCC into development kits for specific systems with their peripherals have a large base of hand-written assembler code.

We looked into three aspects of the calling convention: Use of registers for passing parameters, use of registers for passing return values, cleanup of stack parameters by caller vs. callee. By evaluating a large number of calling conventions, for nine architectures (STM8, Z80, Z180, Z80N, Rabbit 2000, Rabbit 2000A, Rabbit 3000A, TLCS-90, SM83) targeted by SDCC, we found potential for substantial improvement. For five of the architectures our research led to SDCC changing the default calling convention. The new ABI will be the default in the upcoming SDCC 4.2.0 release. For the other four architectures, such a change is still under discussion.

\section{STM8}

\begin{figure}
\centerline{\includegraphics[scale=1.0]{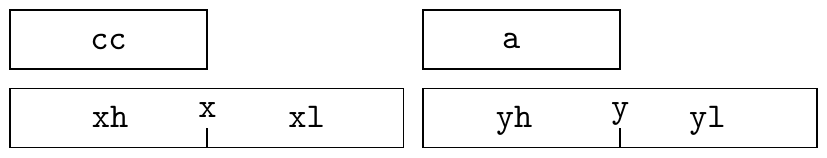}}
\caption{\label{stm8-registers}STM8 registers relevant to the calling convention (\texttt{cc} is the flag register; access to 8-bit parts \texttt{xl}, \texttt{xh} of \texttt{x} and \texttt{yl}, \texttt{yh} of \texttt{y} is via load and store instructions only)}
\end{figure}

The STM8 is a relatively recent and common µC architecture. It has an 8-bit accumulator \texttt{a} and two 16-bit registers \texttt{x} and \texttt{y} (Figure~\ref{stm8-registers}). Most instructions have one register operand and one memory operand. While most instructions available for \texttt{x} are also available for \texttt{y}, many of them are 1 byte longer for \texttt{y}. Besides SDCC, which does not use pseudoregisters, there are competing Raisonance, Cosmic and IAR compilers targeting this architecture.

The calling convention used by SDCC so far passes all parameters on the stack, passes 8-bit return values in \texttt{a}, 16-bit return values in \texttt{x}, 24-bit and 32-bit return values in \texttt{x} and \texttt{y}. Stack parameters are always cleaned up by the caller. This convention was chosen, since it is simple and works for all functions, including those with variable arguments. The other compilers pass 8-bit return values in \texttt{a}, 16-bit return values in \texttt{x}, and 32-bit return values in pseudoregisters. They, too have the caller clean up stack parameters. For functions without variable arguments, Raisonance passes the first argument in \texttt{a}, if it has 8 bits, in \texttt{x} if it has 16 bits, if the first is in \texttt{a}, and the second has 16 bits, it is passed in \texttt{x}, if the first is in \texttt{x}, and the second has 8 bits, it is passed in \texttt{a}, while further arguments are passed on the stack. For functions without variable arguments, Cosmic passes the first argument in \texttt{a}, if it has 8 bits, in \texttt{x} if it has 16 bits, while further arguments are passed on the stack. For functions without variable arguments, IAR passes the first argument that has 8 bits in \texttt{a}, the first argument that has 16 bits in \texttt{x}, while pseudoregisters are used for further and larger arguments.

Users of the STM8 often try to keep their code portable between compilers, and rarely use hand-written assembler code. They tend to be highly sensitive to code size and speed. This makes a good environment for a change in the ABI motivated by reducing overhead.

\section{Z80 and related}

\begin{figure}
\centerline{\includegraphics[scale=1.0]{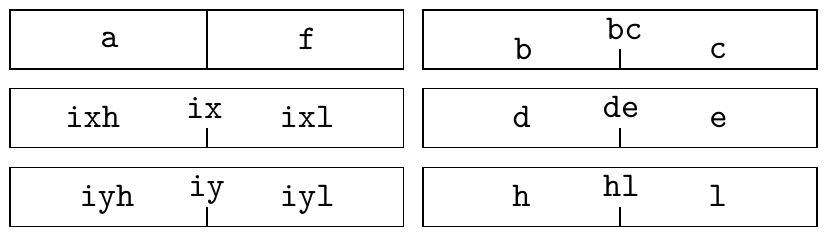}}
\caption{\label{z80-registers}Z80 registers relevant to the calling convention (\texttt{f} is the flag register; \texttt{ix} and \texttt{iy} are not available on SM83; access to 8-bit parts \texttt{ixl}, \texttt{ixh} of \texttt{ix} and \texttt{iyl}, \texttt{iyh} of \texttt{iy} is available on eZ80 only}
\end{figure}

The Z80 is an old processor architecture found in many older computer and video game systems. It has an 8-bit accumulator \texttt{a}, 8-bit registers \texttt{b}, \texttt{c}, \texttt{d}, \texttt{e}, \texttt{h}, \texttt{l} which can also be used as 16-bit registers \texttt{bc}, \texttt{de}, \texttt{hl} and 16-bit registers \texttt{ix}, \texttt{iy} (Figure~\ref{z80-registers}). Most instructions have \texttt{a} as one operand, while the other operand is in another 8-bit register or in memory pointed to by \texttt{hl}, \texttt{ix} or \texttt{iy}. Derived architectures with the same register set, but added instructions include the Z180 and Z80N, which only have a few additional instructions and the eZ80, Rabbit 2000, 3000, 3000A and TLCS-90 which differ substantially in their instruction set from the original Z80. In particular, the eZ80, Rabbit and TLCS-90 have additional 16-bit load instructions. Another architecture derived from the Z80 is the SM83, which was once common in Japanese home appliances such air conditioners and television remotes, and is commonly known for being the architecture of the CPU used in the Game Boy video game system. Compared to the Z80, it lacks the registers \texttt{ix} and \texttt{iy}, as well as many instructions, but it also has a few instructions not present in the original Z80. Competing compilers for these architectures existed historically, but have become less relevant.

The calling convention used by SDCC so far passed all parameters on the stack. For SM83, it passed 8-bit return values in \texttt{e}, 16-bit return values in \texttt{de}, 32-bit return values in \texttt{hlde}. For the other Z80-related targets it passed 8-bit return values in \texttt{l}, 16-bit return values in \texttt{hl}, 32-bit return values in \texttt{dehl}. Stack parameters were always cleaned up by the caller. The original rationale for the choice of the calling convention for Z80 (the oldest version of SDCC we found, SDCC 2.2.0 from 20 years ago already supports the Z80, and the calling convention hasn't changed since; we have been unable to contact the port maintainer from back then) and SM83 is no longer known, but since they are different one can assume that it was a deliberate decision. For the other architectures, SDCC just reused the convention from the Z80. The other compilers used a variety of different calling conventions.

Some of the users of these architectures are highly sensitive to code size or speed. On the other hand, many also have a substantial body of hand-written assembler code. There are big downstream projects (in particular Z88DK and GBDK), that bundle SDCC with further tools and hand-written assembler libraries.

\section{Analysis}

Before looking into calling conventions we wanted to know the common use cases that potentially have a big impact on the compiled programs.

In C, the calling convention can vary depending on the type of the function. Typically, calling conventions take into account only some aspects, such as the return type and argument types width in bit and them being integers vs. pointers vs. floating-point. To get some quantitative data on the relevance of different function types, we created a version of SDCC that outputs data on the function types and calls to them. Since we wanted data that is relevant to code size and speed, we analyzed the intermediate code at register allocation time: function calls that are inlined do not show up in the data, but calls to helper functions used by the compiler (e.g. for software-implemented floating-point operations on hardware without floating-point support) do. A minor drawback of this approach is that we cannot distinguish between types that behave identically, but are different types in C. In particular, C \texttt{char} behaves the same as either \texttt{unsigned char} or \texttt{signed char}, but is a different type.

From analyzing various benchmarks and the standard library, we found that the most commonly called function types are \texttt{float function (float, float)} (very common in floating-point support functions), \texttt{int function (char *, ...)} (e.g. \texttt{printf} from the standard library), \texttt{int function (int, int)} (various functions both in user code and the standard library), \texttt{bool function (float, float)} (common in floating-point support functions), \texttt{int function (char *, char *, unsigned-int)}, \texttt{int function (char *)} (common in string handling, e.g. \texttt{strlen} from the standard library). On the other hand, the most common function types are \texttt{void function (void)}, \texttt{int function (int)} and \texttt{float function (float)}.

\section{Experiments}

To easily evaluate various calling conventions, we created a branch of SDCC that uses nearly no assembler code in the standard library by replacing hand-written assembler functions with generic C code. We also introduced infrastructure that makes it easy to change the calling convention on a per-function basis. We evaluated a large number of calling conventions for each architecture, considering three aspects: Choice of registers for the return value depending on the size of the return value, choice of registers and stack for parameters, choice of caller vs. callee cleanup of stack parameters depending on the size of the return value and on the type of the first parameter. To ensure that the results were not over-optimized to a few common functions, we also repeated the experiments choosing the calling convention for the most commonly called function types independently from the rest. We compiled four benchmarks suitable for execution on the small systems targeted by SDCC (Whetstone~1.2~\cite{Whetstone}, Dhrystone~2.1~\cite{Dhrystone2}, Coremark~1.0~\cite{Coremark}, stdcbench~0.7~\cite{stdcbench}) for each calling convention. For the calling conventions that gave the best results in these experiments, we then created a branch of SDCC that has all standard library functions that were implemented in assembler rewritten to match the new calling convention. For each of STM8, Z80, SM83, Rabbit 3000A, eZ80 and TLCS-90 we evaluated a few thousand different calling conventions that way. All compilation was done using the default optimization goal and strong optimization (using the same options as are used for the compilation of the standard library that comes with SDCC). The default optimization goal tends to favor code size over speed, (though it doesn't go as far as the compiler option \texttt{--opt-code-size} does).

\begin{figure}
\centerline{\includegraphics[scale=1.2]{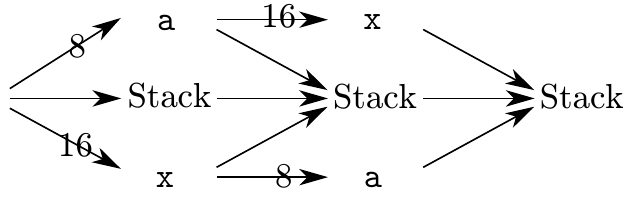}}
\caption{\label{stm8-arguments}STM8 arguments}
\end{figure}

For STM8, we considered any 8-bit register (\texttt{a}, upper and lower halves of {x} and {y}) for 8-bit return values, any 16-bit register for 16-bit return values, any order of the 2 16-bit registers for 32-bit return values. We considered any 8-bit register for 8-bit arguments, and any 16-bit register for 16-bit arguments. Early, it became clear that code generation becomes quite complicated, and has high overhead for calls through function pointers, if there is no free 16-bit register at the time of the call; the experiments also showed that having 16-bit arguments in \texttt{x} provided a substantial advantage, so \texttt{y} would take the role of that free register; this also harmonizes well with the use of \texttt{y} as a frame pointer by SDCC in functions that have more than 256 Bytes of local variables.
For return values, the results showed that the previous convention is good; while some other choices of registers matched the code size and speed of the old one, none surpassed it. For arguments, the convention that worked best turned out to be the one already used by the competing Raisonance compiler: Pass the first argument in \texttt{a}, if it has 8 bits, in \texttt{x} if it has 16 bits, if the first is in \texttt{a}, and the second has 16 bits, it is passed in \texttt{x}, if the first is in \texttt{x}, and the second has 8 bits, it is passed in \texttt{a}, while further arguments are passed on the stack (Figure~\ref{stm8-arguments}). These choices work well for both code size and speed. On the other hand, when it comes to the cleanup of stack parameters by caller vs. callee, there is a trade-off: in general, caller cleanup is faster, while callee cleanup gives smaller code. Doing the cleanup in the caller requires extra effort in code size and speed, but the size overhead happens only once as opposed to every call site; this overhead is bigger if there are few free registers at the end of the call, i.e. for functions with large return values. Also, callee cleanup can hinder tail call optimization. The use of callee cleanup for functions that return \texttt{void} or a value of up to 16 bits (so that one 16-bit register is still free), and for functions where both the return value and the first parameter are of type \texttt{float} looks like a good choice: for the former functions the overhead is small, for the latter functions, due to the STM8 not having hardware floating-point support, the extra runtime is rather small relatively to the effort required by floating-point computations inside the function.

\begin{figure*}
\centerline{
\subcaptionbox{\label{results:size-stm8}STM8 code size}{\includegraphics{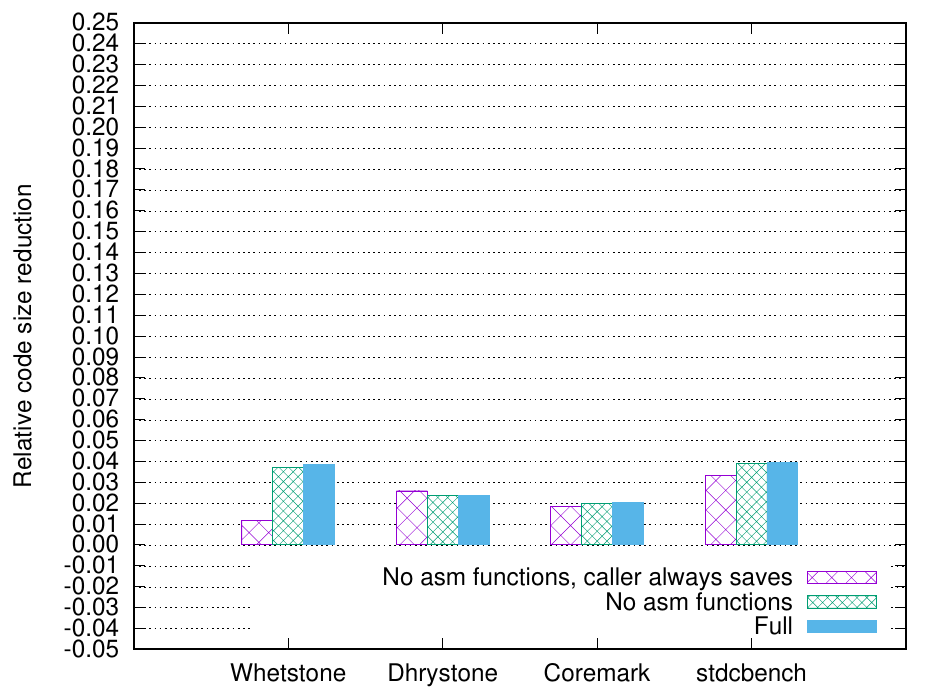}}
\subcaptionbox{\label{results:score-stm8}STM8 score}{\includegraphics{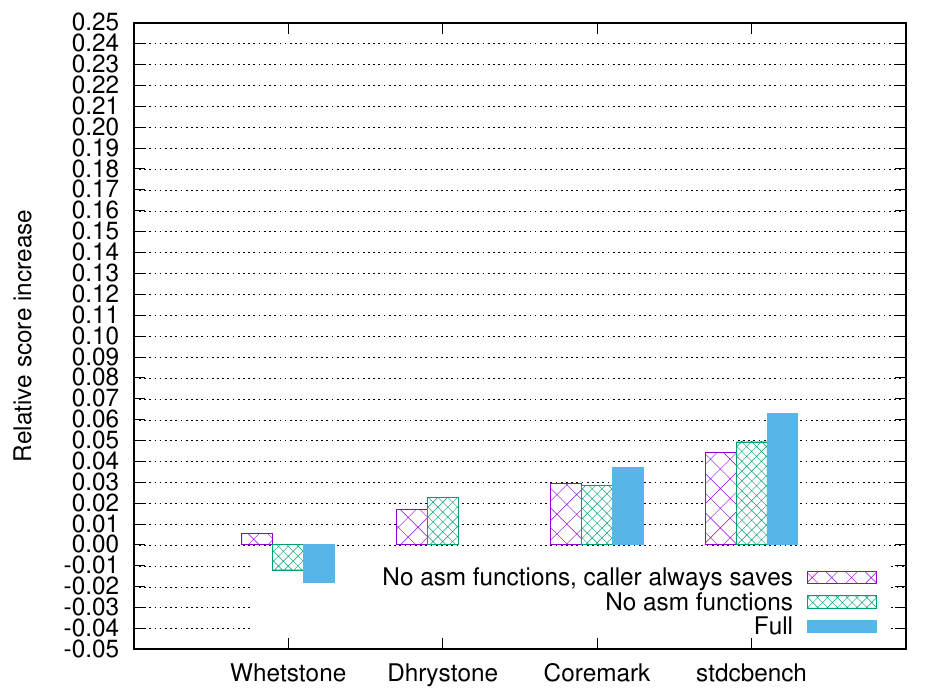}}
}
\centerline{
\subcaptionbox{\label{results:size-z80}Z80 code size}{\includegraphics{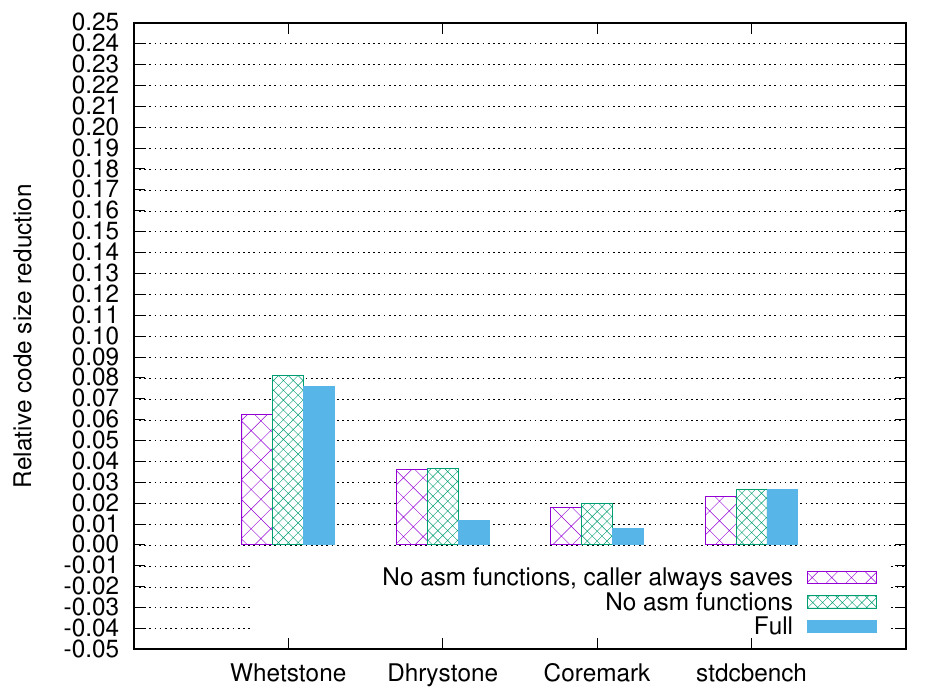}}
\subcaptionbox{\label{results:score-z80}Z80 score}{\includegraphics{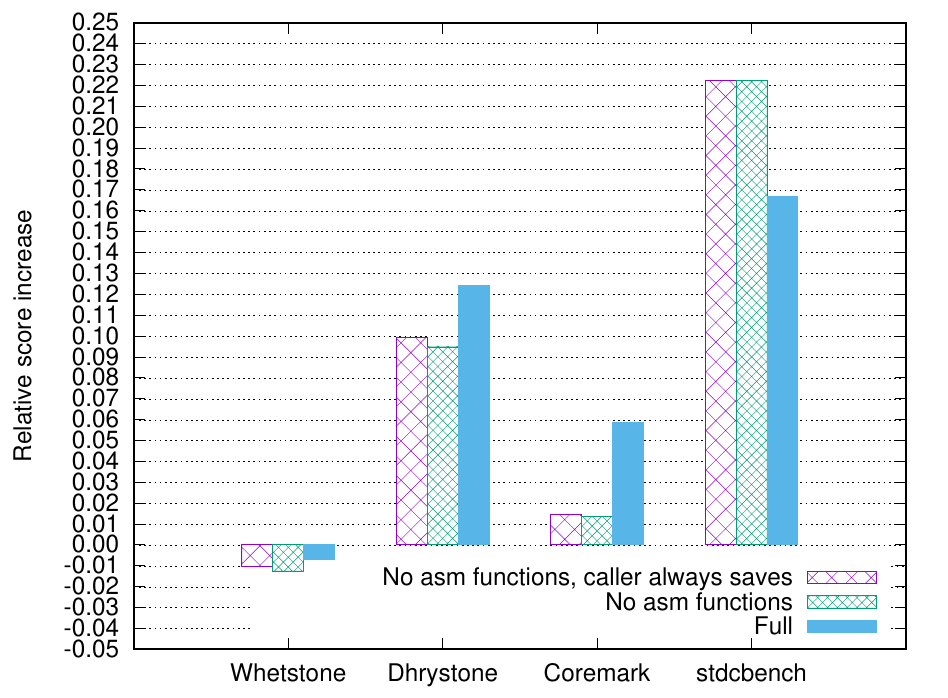}}
}
\centerline{
\subcaptionbox{\label{results:size-r3ka}Rabbit 3000A code size}{\includegraphics{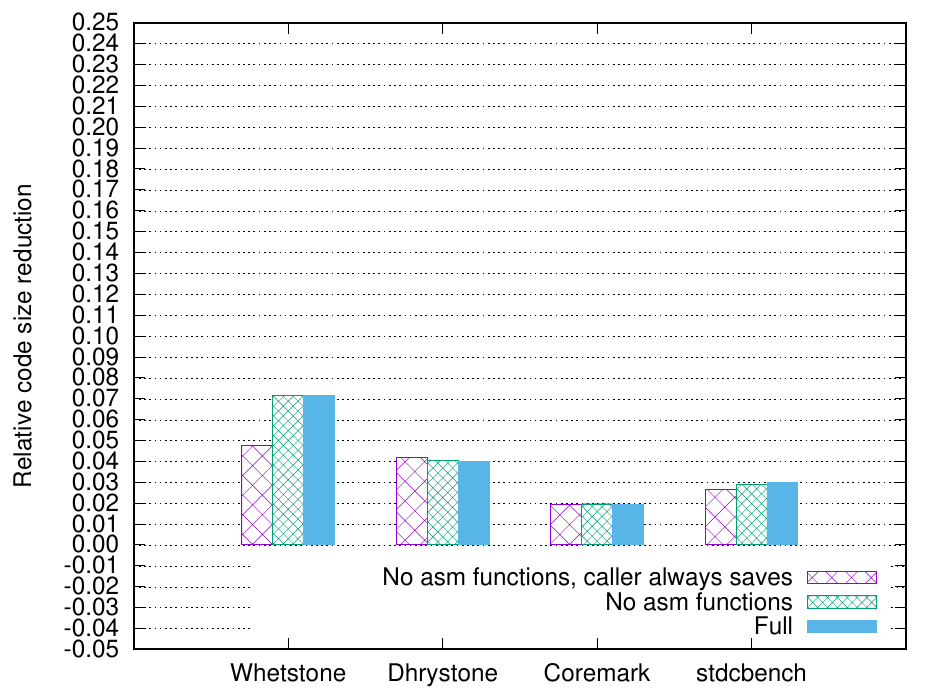}}
\subcaptionbox{\label{results:score-r3ka}Rabbit 3000A score}{\includegraphics{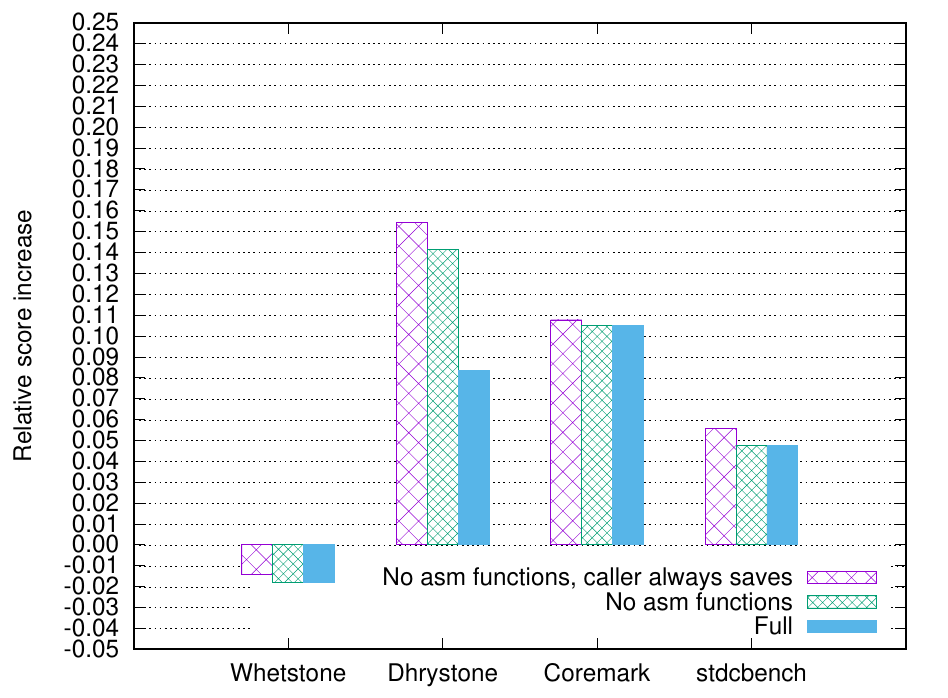}}
}
\caption{\label{results}Experimental results: Code size and speed impact of changing the calling convention}
\end{figure*}

We compiled the benchmarks for the STM8AF board from the STM8A-Discovery kit. The main results can be seen in Figure~\ref{results:size-stm8} for code size and Figure~\ref{results:score-stm8} for speed. For each benchmark, the left bar compares a variant of the new convention that has the caller always restore the stack, to the old convention in the SDCC variant without assembler-imple\-mented library functions. The middle bar compares the new convention to the old convention in the SDCC variant without assembler-imple\-mented library functions. The right bar compares the new convention to the old convention as it will be in the next SDCC release (i.e. both with the assembler-imple\-mented library functions). In Whetstone, having the callee clean up stack parameters saves code size at the cost of code speed. In summary, we see that the change of the calling convention brings substantial improvements in both code size and speed, which is not surprising given the ad-hoc way the old calling convention had been chosen.

For Z80, SM83, Rabbit 2000, 2000A and 3000A, eZ80 and TLCS-90, we considered any 8-bit register for 8-bit return values, any 16-bit register other than \texttt{ix} and \texttt{iy} for 16-bit return values, any order of 16-bit registers other than \texttt{ix} and \texttt{iy} for 32-bit return values. We considered any 8-bit register for a 8-bit arguments, and any 16-bit register other than \texttt{ix} and \texttt{iy} for 16-bit arguments, any order of 16-bit registers other than \texttt{ix} and \texttt{iy} for 32-bit arguments. We excluded \texttt{ix} and \texttt{iy} since some systems reserve their use for a BIOS or OS, and SDCC thus needs to be able to generate code that does not use them when requested by command-line options.

\begin{figure}
\centerline{\includegraphics[scale=1.2]{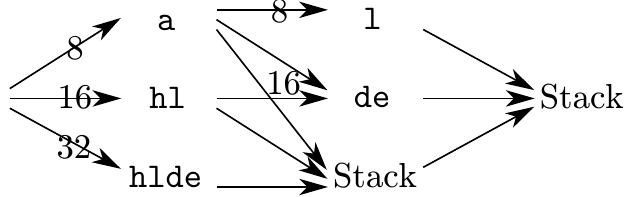}}
\caption{\label{z80-arguments}Z80 / Z80N / Z180 arguments}
\end{figure}

\begin{figure}
\centerline{\includegraphics[scale=1.2]{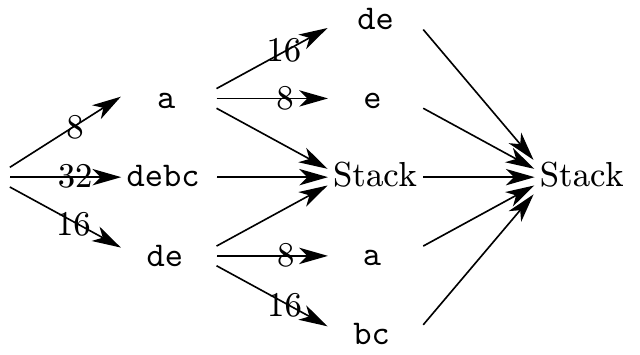}}
\caption{\label{sm83-arguments}SM83 arguments}
\end{figure}

\begin{figure}
\centerline{\includegraphics[scale=1.2]{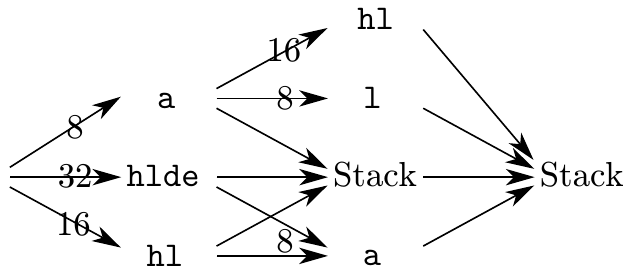}}
\caption{\label{r3ka-arguments}Rabbit 3000A / eZ80 / TLCS-90 arguments}
\end{figure}

Surprisingly, we found that the best calling conventions differed a lot from the previously used ones.
For Z80, we found that having 8-bit return values in \texttt{a}, 16-bit return values in \texttt{de}, 32-bit return values in \texttt{hlde} worked best, a calling convention very close to the one SDCC used for SM83. For the arguments, the best choice was to have the first argument in \texttt{a} if it has 8 bits, in \texttt{hl} if it has 16 bits, in \texttt{hlde} if it has 32 bits. If the first argument is in \texttt{a}, and the second has 8 bits, it is passed in \texttt{l}. If the first argument is in \texttt{a} or \texttt{hl}, and the second has 16 bits, it is passed in \texttt{de}. Further arguments are passed on the stack (Figure~\ref{z80-arguments}). Regarding caller vs callee cleanup the same choice as for STM8 works well.
For SM83, we found that having 8-bit return values in \texttt{a}, 16-bit return values in \texttt{bc}, 32-bit return values in \texttt{debc} worked best. For the arguments, the best choice was to have the first argument in \texttt{a} if it has 8 bits, in \texttt{de} if it has 16 bits, in \texttt{debc} if it has 32 bits. If the first argument is in \texttt{a}, and the second has 8 bits, it is passed in \texttt{e}, if the second has 16 bits, it is passed in \texttt{de}. If the first argument is in \texttt{de}, and the second has 8 bits, it is passed in \texttt{a}, if the second has 16 bits, it is passed in \texttt{bc} (Figure~\ref{sm83-arguments}). For SM83, callee cleanup is a good choice for all functions (the register pair \texttt{hl} is not used for the return value, and thus free at the function end, which together with the SM83 \texttt{add sp, \#d} stack-pointer adjustment instruction allows to generate efficient code for callee cleanup).
For Rabbit 2000, 2000A and 3000A, eZ80 and TLCS-90, we found that having 8-bit return values in \texttt{a}, 16-bit return values in \texttt{hl}, 32-bit return values in \texttt{hlde} worked best. For the arguments, the best choice was to have the first argument in \texttt{a} if it has 8 bits, in \texttt{hl} if it has 16 bits, in \texttt{hlde} if it has 32 bits. If the first argument is in \texttt{a}, and the second has 8 bits, it is passed in \texttt{l}. If the first argument is in \texttt{a}, and the second has 16 bits, it is passed in \texttt{hl}. If the first argument is in \texttt{hl} or \texttt{hlde}, and the second has 8 bits, it is passed in \texttt{a} (Figure~\ref{r3ka-arguments}).

We compiled the benchmarks for the Z80-MBC2 and RCM3319, single-board computers using the Z80 and Rabbit 3000A. The main results can be seen in Figures~\ref{results:size-z80} and \ref{results:size-r3ka} for code size and Figures~\ref{results:score-z80} and \ref{results:score-r3ka} for speed. The meanings of the various bars are the same as for STM8 above (for the Rabbit 3000A, there has not yet been a final decision by the SDCC project if the calling conventions presented here will be the ones used in the future). Again, we see a substantial improvement in code size and speed, which here is more surprising than for STM8. However, for Whetstone, the new convention gives us a speed regression, even when the caller cleans up the stack parameters. This is apparently due to the register allocator having a bit more freedom regarding the register parameters, and the freedom is used for optimizations for code size at the cost of some speed.

We expect that over time, the advantages for the new convention will become even a bit stronger: the peephole optimizer, an optimization stage after code generation, had its current ruleset written when there was only the old calling convention, so there is potential for improvement taking into account the code that code generation commonly generates for the new convention.

A question that had delayed consensus on the future calling convention for Z80, Z180, Z80N, Rabbit 2000, Rabbit 2000A, Rabbit 3000A, eZ80 and TLCS-90 which has since been resolved, was that of 8-bit parameters on the stack being passed as 8-bit vs. 16-bit values. Since the Z80, and related have 16-bit push instructions only, on the caller side a 16-bit push saves code size and speed, but it comes at the cost of stack space. This was thus perceived as a question of code size and speed on one hand versus stack space on the other hand. However, further experiments we did recently resolved this showing that practically, passing these as 8-bit values tends to be better even for code size and speed: having the parameters closer to the current stack pointer allows some extra optimizations on stack accesses in the callee, the peephole optimizer can sometimes use a single 16-bit push to pass two 8-bit arguments, and stack cleanup can be a bit more efficient. SDCC then changed the default calling convention for Z80, Z80N and Z180 to the one we proposed.

\section{Future Work}

While we looked into efficient calling conventions for multiple architectures, there are more such architectures, for which similar experiments could be done, both in SDCC and in other compilers. The latter could also include looking at compilers for languages other than C.

We also expect that our experiments should be redone in about 10 years, as we expect future ISO C standards to change the way C is used, including what is required in calling conventions, as future standards are likely to introduce features that require support in the calling convention (e.g. lambda, closures, integer types of arbitrary width in bits, a separate error reporting channel). Choosing an efficient convention for these new features will require data from their use, and thus some years of user experience with the new features.

SDCC rather quickly changed the default calling convention for the STM8 and SM83. After further discussion, the change was also made for Z80, Z80N and Z180. While our research showed clear advantages to such a change, there is also a cost to breaking binary compability. A remaining open question is if the advantage in code size and speed of having different calling conventions for Z80, Z180, Z80N on one side vs. Rabbit 2000, Rabbit 2000A, Rabbit 3000A, eZ80, TLCS-90 on the other is worth the extra maintenance burden on maintainers of cross-platform libraries written in assembler (and on SDCC developers) of having to learn two different calling conventions: Just using the new convention now used for Z80 is already a lot better than the old, also when used for the Rabbit 2000, 2000A, 3000A, eZ80 and TLCS-90, though not as good as the one we propose for these architectures.

\section{Conclusion}

For irregular architectures, the choice of a calling convention, including which registers to use for return values and parameters makes a big difference in calling convention overhead. Even for established compilers, the potential for further improvement can be substantial enough to make it worth breaking the ABI by changing the calling convention.

\bibliographystyle{plain}
\bibliography{callingconvention}

\end{document}